  \documentstyle[12pt,procsla,epsf,epsfig]{article}
 
  \catcode`\@=11
  \long\def\@makefntext#1{
  \protect\noindent \hbox to 3.2pt {\hskip-.9pt
  $^{{\ninerm\@thefnmark}}$\hfil}#1\hfill}              
 
  \def\@makefnmark{\hbox to 0pt{$^{\@thefnmark}$\hss}}  
 
  \def\ps@myheadings{\let\@mkboth\@gobbletwo
  \def\@oddhead{\hbox{}
  \rightmark\hfil\ninerm\thepage}
  \def\@oddfoot{}\def\@evenhead{\ninerm\thepage\hfil
  \leftmark\hbox{}}\def\@evenfoot{}
  \def\sectionmark##1{}\def\subsectionmark##1{}}
 
\begin{document}
\pagestyle{empty}
\begin{flushright}
{CERN-TH/97-10}
\end{flushright}
\vspace*{10mm}
\begin{center}
{\bf BERTOTTI--ROBINSON GEOMETRY AND SUPERSYMMETRY} \\
\vspace*{2cm} 
{\bf S. Ferrara}$^{*)}$ \\
\vspace{0.3cm}
Theoretical Physics Division, CERN \\
CH - 1211 Geneva 23 \\
\vspace*{2cm}  
{\bf ABSTRACT} \\ \end{center}
\vspace*{5mm}
\noindent
The role of Bertotti--Robinson geometry in the attractor mechanism of extremal black
holes is described for the case of $N = 2$ supersymmetry.  Its implication
for a model-independent derivation of the Bekenstein--Hawking entropy formula
is discussed.
 
\vspace*{3cm} 
\noindent

\begin{center}
{\it Talk given at the 12th Italian Conference on}\\
{\it General Relativity and Gravitational Physics}\\
{\it Rome, September 1996}
\end{center}
\vspace*{4.5cm}
\begin{flushleft} CERN-TH/97-10 \\
January 1997
\end{flushleft}
\vfill\eject

\setcounter{page}{1}
\pagestyle{plain}
\newpage
  \centerline{\normalsize\bf BERTOTTI--ROBINSON GEOMETRY AND SUPERSYMMETRY}
  \baselineskip=22pt
  \baselineskip=16pt

  \vspace*{0.6cm}
  \centerline{\footnotesize Sergio FERRARA}
  \baselineskip=13pt
  \centerline{\footnotesize\it Theory Division, CERN}
  \baselineskip=12pt
  \centerline{\footnotesize\it 1211 Geneva 23, Switzerland}
  \centerline{\footnotesize E-mail: ferraras@vxcern.cern.ch}
  \vspace*{0.3cm}
  \vspace*{0.3cm}
  \baselineskip=13pt
 
  \vspace*{0.9cm}
  \abstracts{The role of Bertotti--Robinson geometry in the attractor mechanism of extremal black
holes is described for the case of $N = 2$ supersymmetry.  Its implication
for a model-independent derivation of the Bekenstein--Hawking entropy formula
is discussed.}
 
  \normalsize\baselineskip=15pt
  \setcounter{footnote}{0}
  \renewcommand{\thefootnote}{\alph{footnote}}
 \section{Extremal Black Holes and Attractors}
In this report, I will discuss some recent work on the macroscopic determination of the
Bekenstein--Hawking entropy-area formula for extremal black holes, using duality symmetries of
effective string theories encoded in the supergravity low-energy actions.  A new dynamical
principle emphasizes the role played by ``fixed scalars" and Bertotti--Robinson type geometries in
this determination.
Supersymmetry seems to be   related to dynamical systems with
fixed points describing  the equilibrium and
stability\footnote {A point  $ x_{\rm fix}$  where the phase
velocity $v(x_{\rm fix})$ is vanishing is named a {\it fixed
point} and represents the system in equilibrium,
$v(x_{\rm fix})=0 $. The fixed point is said to be an {\it
attractor}  of some motion 
$x(t)$ if 
$
\lim_{t  \rightarrow \infty} x(t) = x_{\rm fix}(t).
$}. The particular property of  the long-range behavior of
dynamical flows in dissipative systems is the following: in
approaching the attractors  the orbits  lose practically all
memory of their initial conditions, even though the dynamics
is strictly deterministic. 

The first known example of such attractor behavior in
the supersymmetric system was discovered in the context of
$N=2$  extremal black holes\cite{FKS,S}. The corresponding
motion describes the behavior  of the moduli fields as they
approach the core of the black hole. They evolve according to
a damped geodesic equation  (see eq. (20) in\cite{FKS}) until
they run into the fixed point near the black hole horizon. The
moduli at fixed points were shown to be given  as ratios of
charges in the pure magnetic case\cite{FKS}. It was further shown 
that this phenomenon
extends to the generic case when both electric and magnetic charges are
present\cite{S}. The inverse distance to the horizon plays
the role of the evolution parameter in the corresponding
attractor. By the time moduli  reach the horizon they lose
completely the information about the initial conditions, i.e.
about  their values far away from the black hole, which
correspond to the values of various coupling constants, see
Fig.~1. 
\begin{figure}
\hglue2.5cm
\epsfig{figure=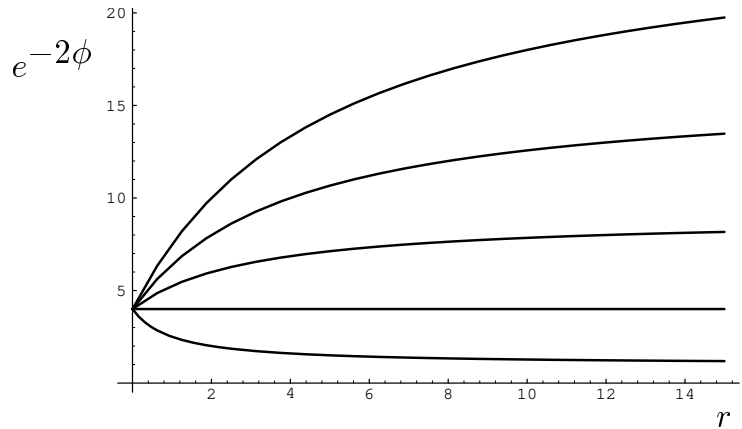,width=10cm}
\caption[]{Evolution of the dilaton from various initial conditions at
infinity to a common fixed point at $r = 0$.}
\end{figure}
 The recent result reported
here\cite{feka} is the derivation of the universal property of  the stable
fixed point of the supersymmetric attractors: fixed point   is defined by
the new {\it principle of a  minimal central charge}\footnote{We are
assuming that the extremum is a minimum, as it can be
explicitly verified in some models. However for the time being
we cannot exclude situations with different extrema or even
where the equation $D_i Z =0$ has no solutions.}~and the area
of the horizon  is proportional to   the square of the central
charge, computed at the point where it is extremized in the
moduli space. In $N=2$,  $d=4$ theories, which is the main object
of our discussion here,  the extremization has to be
performed in the moduli space of the special geometry and  is
illustrated in  Fig. 1. This results in the following formula
for the Bekenstein--Hawking entropy $S$, which is proportional
to the quarter of the area of the horizon:
\begin{equation} S={A\over 4}  =  \pi |Z_{\rm fix}|^2  \ ,
\qquad d=4 \ .
\end{equation}

This result allows generalization for higher dimensions, for
example, in five-dimensional space-time one has
\begin{equation} S={A\over 4}  \sim   |Z_{\rm fix}|^{3/2}    \
, \qquad d=5 \ .
 \end{equation}
  
There exists a  beautiful phenomenon in the black hole
physics: according to the no-hair theorem, there is a limited
number of parameters\footnote{This number can be quite large,
e.g. for $N=8$ supersymmetry one can have 56 charges and 70
moduli.}~which describe space  and physical fields far away
from the black hole. In application to the  recently studied
black holes in string theory,   these parameters include the
mass, the electric and magnetic charges, and the asymptotic
values of the scalar fields. 

 It appears that  for supersymmetric black holes one can prove
a new, stronger version of the no-hair theorem:    black holes
lose all their scalar hair near the horizon. Black hole
solutions near the horizon are characterized only by those
discrete parameters which correspond to conserved charges
associated with gauge symmetries, but not by the values of the
scalar fields at infinity which may change continuously.

A   simple example of this attractor mechanism is given by the
dilatonic black holes of the heterotic string theory
\cite{G,US}. The modulus of the
central charge in question  which is equal to the ADM mass  is
given by the formula
\begin{equation} M_{ADM} =|Z| =  {1\over 2} 
({\mbox{e}}^{-\phi_0}  |p| +  {\mbox{e}}^{\phi_0}  |q|)
\ .
\label{rec}\end{equation} In application to this case the
general theory, developed in this paper gives the following 
recipe to get the area

i) Find the extremum of the  modulus of the central charge as
a function of a dilaton
${\mbox{e}}^{2\phi_0}=g^2$ at fixed charges
\begin{equation} {\partial \over \partial g} |Z| (g, p,q) =  
{1\over 2} {\partial \over \partial g}\left( \frac{ 1}{g}  |p|
+ g  |q| \right) = -\frac{ 1}{g^2}   |p| +   |q| =0 .
\end{equation}

ii) Get the fixed value of the moduli
\begin{equation} g_{\rm fix}^2=  {\left |p\over q \right |}  \
.
\end{equation}

iii) Insert the fixed value into your  central charge formula
(\ref{rec}), get the fixed value of the  central charge:  the
square of it is proportional to the area of the horizon  and
defines the Bekenstein--Hawking entropy
\begin{equation} S = {A\over 4}  =  \pi |Z_{\rm fix}|^2 =  \pi
|pq| \ .
\end{equation} This indeed coincides with the result obtained
before by completely different methods\cite{US,Entr}.

In general supersymmetric  $ N=2$ black holes have an ADM  mass
$M$ depending on  charges $(p, q)$ as well as on moduli $z$
through the holomorphic symplectic sections  $\left( X^\Lambda
(z), F_\Lambda (z)\right)$\cite{Cer}$^-$\cite{Ser1}. The moduli present
the values of the scalar fields of the theory far away from
the black hole. The general formula for the mass of the state
with one half of unbroken supersymmetry of  $N=2$  supergravity
interacting with vector multiplets as well as with
hypermultiplets is\cite{Cer}$^-$\cite{Ser1}
\begin{equation}\label{F1} M^2=|Z |^2  \ , 
\end{equation} where the central charge is\cite{Cer} 
\begin{equation} Z(z, \bar z, q,p) = e^{K(z, \bar z)\over 2} 
(X^\Lambda(z)  q_\Lambda - F_\Lambda(z)
\, p^\Lambda)= (L^\Lambda q_\Lambda - M_\Lambda p^\Lambda) \ , 
\end{equation}
 so that
\begin{equation}\label{F2} M^2_{ADM} =|Z |^2 = M^2_{ADM}
(z,\bar z, p, q) \ .
\end{equation}  The area, however, is only charge  dependent: 
\begin{equation}\label{F3} A=A(p, q) \ .
\end{equation}  This happens since the values of the moduli
near the horizon are driven to the fixed point  defined by the
ratios of the charges.  This mechanism was explained before in
\cite{FKS} and
\cite{S} on the basis of the conformal gauge formulation of 
$N=2$ theory\cite{WLP}.

This attractor mechanism is by no means an exclusive  property
of only $N=2$ theory in four dimensions. Our analysis suggests
that  it may be a quite  universal phenomenon in any
supersymmetric theory. It has in fact been extended to all $N > 2$ theories in
four dimensions and to all theories in five dimensions\cite{feka2}.  Further possible
extensions to higher dimensions and to higher extended objects ($p$-branes) have
also been discussed in recent literature\cite{andafe}.

In this report we will use the ``coordinate free" formulation
of the special geo-metry\cite{CAFP,Cer,Ser1} which will allow us to present a
symplectic invariant description of the system. We will be
able to show that the unbroken supersymmetry requires the
fixed point of attraction  to be defined by the solution of
the duality symmetric equation 
\begin{equation} D_i Z = (\partial _i + {1\over 2}K_i) Z
(z,\bar z, p, q) =0 \ ,
\end{equation} which implies,  
\begin{equation} {\partial \over \partial z^i}  |Z|=0
\end{equation} at 
\begin{equation} Z=Z_{\rm fix}= Z\left( L^\Lambda(p,q),
M_\Lambda(p,q),p,q\right) \ .
\end{equation}

 Equation $\partial _i  |Z|=0$    exhibits the {\it minimal
area principle} in the sense that the area is defined by the
extremum of the central charge in the moduli space of the
special geometry, see Fig. 2 illustrating this point.  Upon
substitution of this extremal values of the moduli into the
square of the central charge we get the Bekenstein--Hawking
entropy,
\begin{equation} S= {A\over 4} =  \pi |Z_{\rm fix} |^2 \ .  
\end{equation}
\begin{figure}
\hglue2.5cm
\epsfig{figure=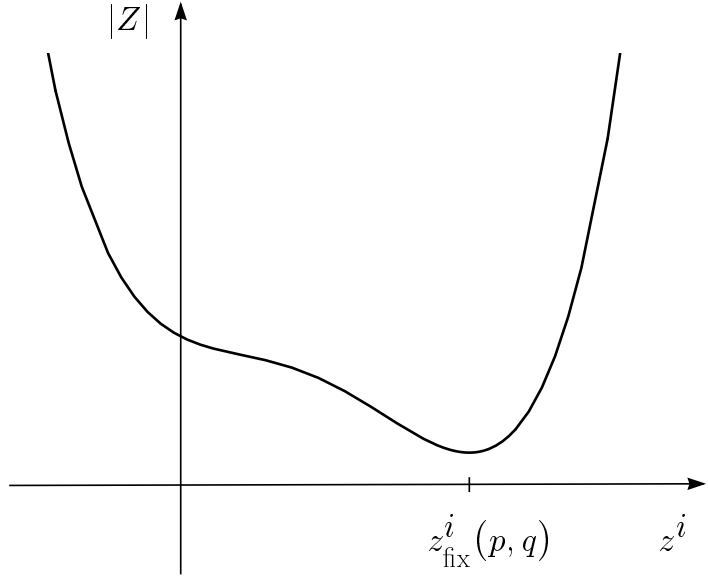,width=10cm}
\caption[]{Extremum of the central charge in the moduli space.}
\end{figure}
 The area of the black hole horizon  has also an
interpretation as the mass of the  Bertotti--Robinson universe
\cite{BR}  describing the near horizon geometry.
\begin{equation} A/4\pi  = M^2_{BR} \ .  
\end{equation} This mass, as different from the ADM mass,
depends only on charges since the moduli near the horizon are
in their fixed point equilibrium positions,
\begin{equation}
 M^2_{BR} =|Z_{\rm fix}|^2 = M^2_{BR} (p,q) \ .
\end{equation} 
Note that in the Einstein--Maxwell system
without scalar fields the ADM mass of the extreme
supersymmetric black hole simply coincides with the
Bertotti--Robinson one, both being functions of charges:
\begin{equation} M^2_{ADM} (p,q) =M^2_{BR} (p,q)\ .  
\end{equation}

  We will describe below a near horizon black holes of $N=2$
supergravity interacting with vector and hyper multiplets. The
basic difference from the pure $N=2$ supergravity solutions
comes from the following: the metric near the horizon is of
the Bertotti--Robinson type, as before. However, the
requirement of unbroken supersymmetry and duality symmetry
forces the moduli to become functions of the ratios of
charges, i.e. take the fixed point values. We will describe
these configurations, show that they provide the restoration
of full unbroken $N=2$ supersymmetry near the horizon. We will
call them $N=2$ attractors, see Sec. 2. We will briefly report on the
extension of the attractor mechanism to more general (higher $N$ and higher
$D$) theories in Section~3.

 \section{Bertotti--Robinson Geometry and fixed scalars} 

The special role of the Bertotti--Robinson metric in the
context of the solitons in supergravity was explained 
 by  Gibbons \cite{Gibb}. He suggested to consider the
Bertotti--Robinson (${\cal BR}$)  metric as an alternative,
maximally supersymmetric, vacuum state. The extreme
Reissner--Nordstr\"om     metric spatially interpolates between
this vacuum and the trivial flat one, as one expects from a
soliton.

Near the horizon all  $N=2$  extremal black holes with one half
of unbroken supersymmetry restore the complete  $N=2$  unbroken
supersymmetry. This phenomenon of the doubling of the
supersymmetry  near the horizon was discovered    in the
Einstein--Maxwell system in\cite{Gibb}.  It was explained in 
\cite{K} that the manifestation of this doubling of unbroken
supersymmetry is the appearance of a covariantly constant on
shell  superfield of $N=2$ supergravity. In presence of a
dilaton this mechanism was studied in 
\cite{KP}. In the context of exact four-dimensional black
holes, string theory and conformal theory on the world-sheet
the 
${\cal BR}$ space-time was studied in\cite{LS}.
 In more general setting the idea of vacuum interpolation in
supergravity via super p-branes was developed in\cite{GT}.

We will show here using the most general supersymmetric system
of $N=2$ supergravity interacting with vector multiplets and
hypermultiplets how this doubling of supersymmetry occurs and
what is the role of  attractors in this picture. The
supersymmetry transformation for the gravitino, for the
gaugino and for the hyperino  are given in the manifestly
symplectic covariant formalism\cite{CAFP,Ser1}
\footnote{The notation is given in\cite{Ser1}.}~in the absence
of fermions and in absence of gauging as follows:
\begin{eqnarray}
\delta \psi_{A\mu} &=& {\cal D}_\mu \epsilon_A +\epsilon_{AB}
T_{\mu\nu}^- \gamma^\nu \epsilon^B \ ,
\nonumber\\
\delta \lambda^{iA} &=& i\gamma^\mu \partial_\mu z^i
\epsilon^A+ {i\over2} {\cal
F}^{i-}_{\mu\nu}\gamma^{\mu\nu}\epsilon_B\epsilon^{AB}\ ,
\nonumber\\  
\delta \zeta_\alpha &=& i \,{\cal U}^{B \beta} _u
\partial_\mu  q^u \gamma^\mu \epsilon ^A
\epsilon_{AB}
 C_{\alpha \beta} \ ,
\label{ricca}
\end{eqnarray} where $\lambda^{iA}$, $\psi_{A\mu}$ are the
chiral gaugino and gravitino fields,
$\zeta_\alpha$ is a hyperino, $\epsilon_A\ ,\ \epsilon^A$ are
the chiral and antichiral supersymmetry parameters
respectively, $\epsilon^{AB}$ is the $SO(2)$ Ricci tensor. The
moduli dependent duality invariant combinations of field
strength 
$T_{\mu\nu}^- , {\cal F}^{i-}_{\mu\nu}$ are defined by eqs. 
(\ref{tnov}), ${\cal U}^{B \beta} _u $ is the quaternionic
vielbein\cite{BW}.

Our goal is to find solutions with unbroken $N=2$ supersymmetry.
The first one is a standard flat vacuum: the metric is flat,
there are no vector fields, and  all scalar fields in the
vector multiplets as well as in the hypermultiplets take
arbitrary constant values: 
\begin{equation} ds^2 = dx^\mu dx^\nu \eta_{\mu\nu}\ , \qquad
T_{\mu\nu}^-={\cal F}^{i-}_{\mu\nu}=0\ ,  \qquad z^i = z^i_0 \
,  \qquad q^u=q^u_0 \ .
\end{equation} This solves the Killing conditions $\delta
\psi_{A\mu}=\delta \lambda^{iA}=\delta
\zeta_\alpha=0$ with constant unconstrained values of the
supersymmetry parameter $\epsilon_A$. The unbroken
supersymmetry manifests itself in the fact that each
non-vanishing scalar field represents the first component of a
covariantly constant $N=2$ superfield for the vector and/or 
hyper multiplet, but the supergravity superfield vanishes.

The second solution with unbroken supersymmetry is much more
sophisticated. First, let us solve the equations for the
gaugino and hyperino by using only a  part of the previous
ansatz: 
\begin{equation} {\cal F}^{i-}_{\mu\nu}=0\ ,  \qquad \partial
_\mu z^i =0\ ,  \qquad \partial _\mu q^u=0
 \ .
\end{equation} The Killing equation for the gravitino is not
gauge invariant. We may therefore consider the variation of
the gravitino field strength the way it was done in\cite{K},
\cite{KP}. Our ansatz for the metric will be to use the geometry
with the vanishing scalar curvature and Weyl tensor and
covariantly constant graviphoton  field strength
$T_{\mu\nu}^-  $ : 
\begin{equation} R= 0\ , \qquad  C_{\mu \nu \lambda \delta}=0
\ ,  \qquad {\cal D}_\lambda  (T^-_{\mu\nu} )=0  \ .
\end{equation} It was explained in\cite{K},\cite{KP} that such
configuration corresponds to a covariantly constant 
superfield of $N=2$ supergravity $W_{\alpha \beta} (x, \theta)$,
whose first component is given by a two-component  graviphoton
field strength $T _{\alpha \beta} $. The doubling of
supersymmetries near the horizon happens  by the following
reason. The algebraic condition for the choice of broken
versus unbroken supersymmetry is given in terms of the
combination of the Weyl tensor plus or minus  a covariant
derivative of the graviphoton field strength, depending on the
sign of the charge. However, near the horizon both the Weyl
curvature and the vector part vanish. Therefore both
supersymmetries are restored and we simply have a covariantly
constant superfield
$W_{\alpha \beta} (x, \theta)$. The new feature of the generic
configurations which include vector and hyper multiplets is
that in addition to a covariantly constant superfield of
supergravity
$W_{\alpha \beta} (x, \theta)$, we have covariantly constant
superfields, whose first component is given by the scalars of
the corresponding multiplets.  However, now as different from
the trivial flat vacuum, which admits any values of the
scalars, we have to satisfy the consistency conditions for our
solution, which requires that the Ricci tensor is defined by
the product of graviphoton field strengths,
\begin{equation} R_{\alpha \beta \alpha' \beta'}^{BR} = T
_{\alpha \beta}  \bar T_{\alpha' \beta'} \ ,
\label{I}\end{equation} and that the vector multiplet vector
field strength vanishes
\begin{equation} {\cal F}^{i-}_{\mu\nu}=0 \ .
\label{II}\end{equation} Before analysing these two
consistency conditions in terms of  symplectic structures of
the theory, let us describe the black hole metric near the
horizon.

The explicit form of the metric is taken as a limit near the
horizon $r=|\vec x | \rightarrow0$ of the black hole metric
\begin{equation} ds^2 = -e^{2U} dt^2 + e^{-2U} d\vec x^2 \ ,
\end{equation} where 
\begin{equation}
\Delta e^{-U} =0 \ .
\end{equation} We choose 
\begin{equation} e^{-2U}= {A\over 4\pi |\vec x |^2}= 
{M_{BR}^2\over  r^2} \ , 
\end{equation} where the Bertotti--Robinson  mass is defined by
the black hole area of the horizon
\begin{equation} M_{BR}^2= {A\over 4\pi} \ . 
\end{equation}

We may show that this metric, which is the Bertotti--Robinson
metric  
\begin{equation} ds^2_{BR} = -  {|\vec x |^2\over M_{BR}^2}
dt^2 +  {M_{BR}^2\over  |\vec x |^2} 
 d\vec x^2 \ ,
\end{equation} is conformally flat in the properly chosen
coordinate system.  In spherically symmetric coordinate system
\begin{equation} ds^2_{BR} = -  {r^2\over M_{BR}^2} dt^2 + 
{M_{BR}^2\over  r^2}
 (dr^2 + r^2 d\Omega) \ .
\end{equation} After the change of variables 
$r=M_{BR}^2/\rho$ and $|\vec x | = M_{BR}^2/|\vec y |$   the
metric becomes obviously conformally flat
\begin{equation} ds^2_{BR} = -  {M_{BR}^2\over \rho^2 } dt^2
+  {M_{BR}^2\over  \rho^2}
 (d\rho^2 + \rho^2 d\Omega)  ={M_{BR}^2\over | {\vec y}| ^2 }
( -dt^2 +   d \vec y^2) \ ,
\end{equation} which is in agreement  with the vanishing of
the Weyl tensor.

Now we are ready to describe our solution in terms of
symplectic structures, as defined in
\cite{Ser1}. The symplectic structure of the equations of
motion comes by defining the ${\rm Sp}(2n_V+2)$ symplectic
(antiselfdual) vector field strength
$({\cal F} ^{-\Lambda},{\cal G}^-_{\Lambda} ) $.

Two symplectic invariant combinations
 of the symplectic field strength vectors are:
\begin{eqnarray} T^-&=& M_\Lambda {\cal F}^{-\Lambda}-
L^{\Lambda } {\cal G}_\Lambda^- \nonumber\\
\nonumber\\ {\cal F} ^{-i}&=&G^{i\bar j} ( D_{\bar j} \bar
M_\Lambda {\cal F}^{-\Lambda} -   D_{\bar j}
\bar L^\Lambda {\cal G}_\Lambda^-) \ .
\label{tnov}
\end{eqnarray} The central charge as well as the covariant
derivative of the central charge are defined as follows: 
\begin{equation} 
 Z = -{1\over 2} \int_{S_2} T^- \ , 
\label{tund}
\end{equation} and  
\begin{equation}  Z_i \equiv D_i Z    =  -{1\over 2}
\int_{S_2} {\cal F}^{+\bar j} G_{i\bar j}\ . 
\label{tdod}
\end{equation} The central charge, as well as its derivative, 
are functions of moduli and electric and magnetic charges. The
objects defined by eqs. (\ref{tnov}) have the physical meaning
of being the (moduli-dependent) vector combinations which
appear in the gravitino and gaugino supersymmetry
transformations respectively.
 In the generic point of the moduli space there are two
symplectic    invariants homogeneous of degree 2 in electric
and magnetic charges\cite{Ser1}:
\begin{eqnarray} I_1&=& |Z|^2 + |D_i Z|^2 \ ,\nonumber\\
\nonumber\\ I_2&=& |Z|^2 - |D_i Z|^2 \ .
\end{eqnarray} Note that 
\begin{eqnarray} I_1&=& I_1(p,q,z,\bar z)=-{1\over 2} P^t
{\cal M}({\cal N}) P\ ,
\nonumber\\ I_2&=&I_2(p,q,z,\bar z)=-{1\over 2} P^t {\cal
M}({\cal F}) P \ . 
\label{F}
\end{eqnarray}

Here $P=(p,q)$ and ${\cal M}({\cal N})$ is the real symplectic
$2n+2 \times 2n+2$ matrix

\begin{equation}
\pmatrix{ A & B \cr C & D \cr }
\end{equation} where 
\begin{eqnarray} A&=& {\rm Im} {\cal N} + {\rm Re} {\cal N}
{\rm Im} {\cal N}^{-1} {\rm Re} {\cal N}
\ , \qquad B =- {\rm Re} {\cal N} \, {\rm Im} {\cal N}^{-1}
\nonumber\\
\nonumber\\ C&=&-{\rm Im} {\cal N}^{-1}  {\rm Re} {\cal N} \
,  \hskip 2.7 cm  D= {\rm Im} {\cal N}^{-1} \ .
\end{eqnarray} The vector kinetic matrix  ${\cal N}$ was
defined in Refs.\cite{Cer}$^-$\cite{Ser1}. The same type of matrix appears in
(\ref{F}) with ${\cal N} \rightarrow{\cal F} =F_{\Lambda
\Sigma}$.  Both ${\cal N}, {\cal F}$ are K\"ahler invariant
functions, which means that they depend only on ratios of
sections, i.e. only on $t^\Lambda,  f_\Lambda$\cite{Cer}$^-$\cite{Ser1}. 

The unbroken supersymmetry of the near horizon black hole
requires the consistency condition (\ref{II}), which is also a
statement about the fixed point for the scalars $z^i(r) $ as
functions of the distance from the horizon $r$
\begin{equation} {\partial \over \partial r} \left (z^i(r)
\right ) =0 \qquad  \Longrightarrow 
\qquad  D_i Z=0 \ .
\label{fixed}\end{equation} Thus the fixed point is defined
due to supersymmetry by the vanishing of the covariant
derivative of the central charge. At this point  the critical
values of moduli  become functions of charges, and two
symplectic invariants become equal to each other:
\begin{equation} I_{1 \, \rm fix}= I_{2\,  \rm fix} =(
|Z|^2)_{D_i Z=0} \equiv |Z_{\rm fix}|^2 \ .
\label{I=II}\end{equation} The way to explicitly compute the
above is by solving in a gauge-invariant fashion eq.
(\ref{fixed}),
yielding:
\begin{equation}\label{} p^\Lambda = i(\bar Z L^\Lambda - Z
\bar L^\Lambda)\ , \qquad q_\Lambda = i(\bar Z M_\Lambda - Z
\bar M_\Lambda) \ .
\end{equation}
From the above equations it is evident that $(p,q)$  determine
the sections up to a (K\"ahler) gauge transformation (which
can be fixed setting $L^0 =e^{K/2}$). Vice versa the fixed
point $t^\Lambda$ can only depend on ratios of charges since
the equations  are homogeneous in p,q.

 The first invariant provides an elegant expression of
$|Z_{\rm fix}|^2$ which only involves the charges and the
vector kinetic matrix at the fixed point ${\cal N}_{\rm fix}
={\cal N}\left ( t^\Lambda_{\rm fix},  \bar t^\Lambda_{\rm
fix} , f_{\Lambda\, {\rm fix}}, \bar f_{\Lambda \,{\rm fix}}
\right)$.
\begin{equation} (I_1)_ {\rm fix} =( |Z|^2 + |D_i Z|^2)_{\rm
fix} = -{1\over 2} P^t {\cal M}({\cal N} _{\rm fix}) P = (
|Z_{\rm fix}|^2 ) \ .
\label{N}\end{equation} Indeed eq. (\ref{N}) can be explicitly
verified by using eq. (\ref{old}). For  magnetic solutions
the   area  formula  was derived in\cite{FKS}. This formula
presents the area as the function of the zero component of the
magnetic charge and of the K\"ahler potential at the fixed
point\footnote{In this paper we have a  normalization of
charges which is different from
\cite{FKS} due to the use of the conventions of\cite{Ser1}
and not\cite{WLP}.}.
\begin{equation}\label{old}
 A = \pi (p^0)^2 e^{-K}  \ .
\end{equation} In the symplectic invariant  formalism we may
check that the area formula (\ref{old}) which is valid for the
magnetic solutions (or for generic solutions but in a specific
gauge only) indeed can be brought to the symplectic invariant
form\cite{FKS}:
\begin{equation} A = \pi (p^0)^2 e^{-K} =4 \pi ( |Z|^2 + |D_i
Z|^2)_{\rm fix}  = 4 \pi ( |Z_{\rm fix}|^2 )= -2 \pi 
p^{\Lambda} {\rm Im}  {\cal F}_{\Lambda \Sigma} p^\Sigma \ .  
\end{equation}

One can also check the first consistency condition of unbroken
supersymmetry (\ref{I}), which relates the Ricci tensor to the
graviphoton.  Using the definition of the central charge in
the fixed point we are lead to the formula for the area of the
horizon (which is defined via the mass of the
Bertotti--Robinson geometry) in the following form
\begin{equation}\label{new} M_{BR}^2= {A\over 4\pi} = (
|Z|^2)_{D_i Z=0} \  , \qquad S={A\over 4} = 
 \pi M_{BR}^2 \ .
\end{equation}

The new area formula  (\ref{new}) has various advantages 
following from manifest symplectic  symmetry. It also implies
the  principle of the  minimal mass of the Bertotti--Robinson 
universe, which is given by the extremum in the moduli space
of the special geometry.
\begin{equation}\label{uu2}
\partial _i M_{BR} =0 \ .
\end{equation}

\section{Attractors in more general theories}
The previous analysis admits an extension to higher $N$ theories at $D = 4$ and
to $N \geq 2$ theories at $D = 5$.  The general condition for getting a
non-vanishing entropy-area formula is that extremal black-holes preserve only
one supersymmetry (1/4 for $N = 4$ and 1/8 for $N = 8$).  This condition is
again obtained by extremizing the ADM mass in the moduli space.  In this case the
ADM mass is given, at $D = 4$, by the highest eigenvalue of the central charge
matrix and supersymmetry implies that when the ADM mass is extremized, the other
eigenvalues vanish.  The vanishing of the other eigenvalues imply
that the variation of the spin-1/2 partners of the gravitino vanishes on the
residual unbroken (Killing) supersymmetry\cite{feka2}. 

The entropy-area formula is always given by a $U$-duality invariant expression
built out of the electric and magnetic charges\cite{cvhu}$^,$\cite{ba}.  This is a consequence
of the fact that both for $D = 4$ and 5 the area is given by an appropriate power of the
extremized ADM mass which is a $U$-duality invariant expression (the central
charge is an expression of first degree in terms of electric and magnetic
charges)\cite{feka}$^,$\cite{feka2}.

 Similar arguments show that the attractor mechanism gives a vanishing (or
constant) result for
$D > 5$ since a
$U$-duality invariant expression does not exist in that case.  Similar
considerations can be extended to higher $p$-extended objects in any $D$\cite{andafe}. 

Finally, we note that the mechanism of the doubling of supersymmetry at the
attractor point is still operating in five dimensions\cite{chfegi}.
The analogue of the Bertotti--Robinson geometry (which is $AdS_2 \times S^2$)
is in this case the Tangherlini extremal $D = 5$ black hole \cite{tangh} (with topology of
$AdS_2 \times S^3)$ which indeed admits two Killing spinors.  This is the fixed
moduli geometry of the
$D = 5$ attractors.

Recently many applications of these ideas have been worked out, especially in the
case of string theory compactified on three-dimensional Calabi--Yau complex
manifolds\cite{twentyone}.  Determination of the topological entropy formula by counting
microcospic states in string theory, by means of $D$-brane techniques, has also
been performed\cite{twentytwo} and shown to give results, whenever obtainable, in agreement with
the model-independent determination which uses the attractor mechanism.

Finally it should be mentioned that several properties of ``fixed scalars" have been
investigated\cite{klkr}.  In particular, it has been shown that the attractor mechanism
is also relevant in the discussion of black hole thermodynamics out of
extremality\cite{gikako}.

\section{Acknowledgements}
It was a pleasure to give this talk on the occasion of the 65th birthday of Prof.
Bruno~Bertotti.  The material covered in this talk comes mainly from work done jointly
with Renata~Kallosh and Andrew~Strominger, both of which I would like to thank for very
pleasant and fruitful collaboration.

This work was supported in part by EEC under TMR contract ERBFMRX-CT96-0045, Frascati, and by
DOE grant DE-FG03-91ER40662.

\end{document}